\documentclass[aps, pra,twocolumn, amsmath,amssymb,superscriptaddress
]{revtex4-2}

\usepackage{graphicx} 
\usepackage{hyperref} 
\usepackage{xcolor}   
\usepackage{physics}  
\usepackage{diagbox}  
\usepackage{dcolumn}  
\usepackage{booktabs} 
\usepackage{multirow}
\usepackage{tabularx}
\hypersetup{
    colorlinks=true,
    linkcolor=blue,
    citecolor=blue,
    filecolor=blue,
    urlcolor=blue
}
\usepackage{placeins}

\newcommand{\UMD}{University of Maryland, College Park, Maryland, USA}

\newcommand{\NIST}{Sensor Science Division, National Institute of Standards and Technology, Gaithersburg, Maryland 20899, USA}

\begin{document}


\title{{Spectroscopy of laser cooling transitions in MgF}}

\date{\today}

\author{N. H. Pilgram}
\affiliation{\NIST}

\author{B. Baldwin}
\affiliation{\NIST}
\affiliation{\UMD}
\author{D. S. La Mantia}
\affiliation{\NIST}

 \author{S. P. Eckel}
 \affiliation{\NIST}

\author{E. B. Norrgard}
 \affiliation{\NIST}
 \email{eric.norrgard@nist.gov}

\graphicspath{{./}{./figs/}}

\newcommand{\SPE}[1]{\textcolor{orange}{\textsf{[SPE: #1]}}}
\newcommand{\EBN}[1]{\textcolor{red}{\textsf{[EBN: #1]}}}
\newcommand{\NHP}[1]{\textcolor{blue}{\textsf{[NHP: #1]}}}

\newcommand{\checktext}[1]{{\color{red} #1}}

\begin{abstract}
We measure the complete set of transition frequencies necessary to laser cool and trap MgF molecules.
Specifically, we report the frequency of multiple low $J$ transitions of the $X^2\Sigma^+(v^{\prime\prime}=0,1) \rightarrow A^2\Pi_{1/2}(v^\prime=0)$, $X^2\Sigma^+(v^{\prime\prime}=1,2) \rightarrow A^2\Pi_{1/2}(v^\prime=1)$, and $X^2\Sigma^+(v^{\prime\prime}=1) \rightarrow B^2\Sigma^+(v^{\prime}=0)$ bands of MgF. The measured $X^2\Sigma^+(v^{\prime\prime}=1)\rightarrow B^2\Sigma^+(v^\prime=0)$ spectrum allowed the spin-rotation and hyperfine parameters of the $B^2\Sigma^+(v=0)$ state of MgF to be determined.
Furthermore, we demonstrate optical cycling in MgF by pumping molecules into the $X^2\Sigma^+(v=1,2)$ states.
Optical pumping enhances the spectroscopic signals of transitions originating in the $N^{\prime\prime}=1$ level of the $X^2\Sigma^+(v^{\prime\prime}=1,2)$ states.
\end{abstract}

\maketitle

\section{Introduction}

Laser-cooled and trapped molecules are an emerging technology for quantum computing \cite{Chae2021,Holland2023,Cornish2024}, precision measurement \cite{Hunter2012,Alauze2021,Anderegg2023}, and metrology applications \cite{Norrgard2021, Vilas2023, Manceau2024}.
All laser-cooled and trapped molecules to-date have $^2\Sigma^+$ ground state symmetry.  Following the proposal of Ref.\,\cite{Stuhl2008}, rotational closure is achieved by optical cycling on the $P_1(1)$ branch for  $^2\Sigma^+ \rightarrow \, ^2\Sigma^+$ \cite{Truppe2017} or the $P_1(1)$/$Q_{12}(1)$ branch for   $\,^2\Sigma^+\rightarrow ^2\Pi_{1/2} $ cycling transitions \cite{Barry2014}.  In either of these cases the maximum photon scattering rate is $R_{\rm{sc}}^{\rm{max}} = \Gamma/4$ \cite{Kloter2008,Norrgard2016}, where $\Gamma$ is the radiative decay rate of the excited state. The maximum scattering rate is reduced from this value if rovibrational repump lasers couple to the excited state used for optical cycling (for example, $R_{\rm{sc}}^{\rm{max}} = \Gamma/7$ in the cycling scheme shown in Fig.\,\ref{fig:BS} a).
For alkaline earth monofluorides \cite{Hao2019, Norrgard2023} (as well as many polyatomic molecules with an alkaline earth optical cycling center \cite{Kozyryev2019, Lasner2022, Li2019, Klos2020}), both the $A^2\Pi_{1/2}$ and $B^2\Sigma^+$ states have a more or less diagonal matrix of Franck-Condon factors with the $X^2\Sigma^+$ state.  Thus, either the $A^2\Pi_{1/2}$ or $B^2\Sigma^+$ state can be used for optical cycling with the other used for repumping. 
The $B^2\Sigma^+$ state has been used as the main cycling transition for the laser slowing of CaF \cite{Truppe2016}, as a vibrational repump for the laser cooling of CaOH \cite{Vilas2022}, and has been proposed as the main cycling transitions for the magneto-optical trapping of CaF \cite{Devlin2015} and SrF \cite{Langin2023}.  

It is possible to apply large radiative forces to MgF due to its fast radiative decay rate $\Gamma = 1.316(14)\times10^8$\,s$^{-1}$, low mass $m = 43$\,u, and short cycling transition wavelength $\lambda = 359.3$\,nm.
By utilizing both the $A^2\Pi_{1/2}$ and $B^2\Sigma^+$ for laser cooling in order to optimize the   \linebreak

\onecolumngrid\
\begin{center}\
\begin{figure}[b]\label{fig:BS}\
\includegraphics[scale=.87]{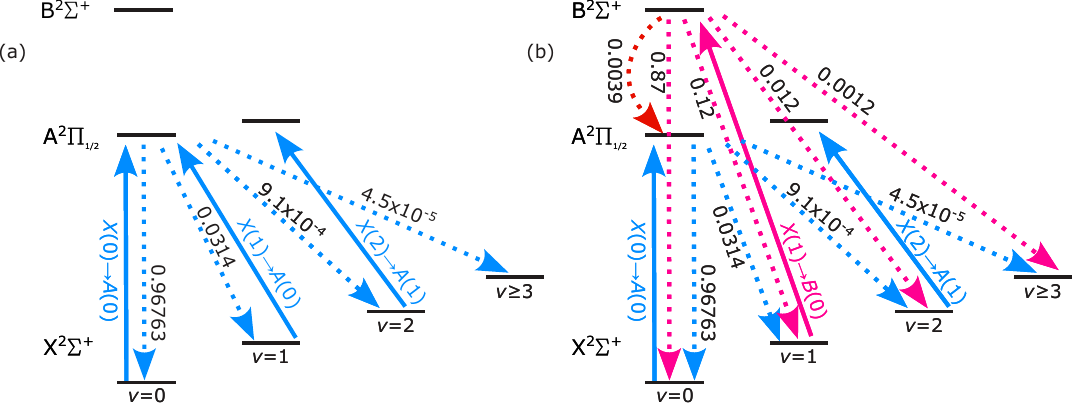}\
\caption{Optical cycling in MgF. Solid arrows depict  $P_1(1)$/$Q_{12}(1)$ $X\rightarrow A$ transitions or $P_1(1)$ $X\rightarrow B$ transitions driven by lasers. Dashed arrows depict decay paths with numbers indicating branching fractions from Ref.\,\cite{Norrgard2023}.  (a) Cycling using only the $A^2\Pi_{1/2}$ excited state.  A $\Lambda$ system is formed by driving both the $X(0,1)\rightarrow A(0)$ transitions. (b) Cycling using  the $B^2\Sigma^+$ excited state for vibrational repumping avoids a $\Lambda$ system. }\
\end{figure}\
\end{center}\
\twocolumngrid\ 

\hfill\linebreak
\noindent radiative force, it should be possible to directly load a MgF magneto-optical trap from a cryogenic buffer gas beam (CBGB) source \cite{Hemmerling2014, Rodriguez2023}.
In Ref.\,\cite{Norrgard2023} we determined  the radiative decay rate and branching fractions of the  $A^2\Pi_{1/2} \rightarrow X^2\Sigma^+$ optical cycling transition in MgF with high precision. 
In that work, we followed the proposal of Ref.\,\cite{NorrgardThesis} to use the $B^2\Sigma^+$ for vibrational repumping (Fig.\,\ref{fig:BS} b) in order to avoid repumping $v^{\prime\prime}=1$ through an excited state in common with the main cooling transition $X^2\Sigma^+(v^{\prime\prime}=0) \rightarrow A^2\Pi_{1/2}(v^\prime=0)$  (such an arrangement of two ground states coupled to a common excited state is often referred to as a ``$\Lambda$ system''\,\cite{Marzoli1994}). 
A detailed understanding of the $B^2\Sigma^+$ state is thus required for laser cooling and trapping of MgF.
However, the previous analysis of the high-temperature, Doppler-limited, spectrum of the $B^2\Sigma^+ - X^2\Sigma^+$ transition of MgF [27] lacked the spectral resolution to detect the fine and hyperfine structure of the $B^2\Sigma^+$ state. 


In this work, we experimentally observe and identify  all transitions necessary for magneto-optical trapping of MgF \cite{Norrgard2023}: main cooling transition \mbox{$X^2\Sigma^+(v^{\prime\prime}=0) \rightarrow A^2\Pi_{1/2}(v^\prime=0)$} as well as vibrational repumping transitions \mbox{$X^2\Sigma^+(v^{\prime\prime}=1) \rightarrow A^2\Pi_{1/2}(v^\prime=0)$}, \mbox{$X^2\Sigma^+(v^{\prime\prime}=2) \rightarrow A^2\Pi_{1/2}(v^\prime=1)$}, and \mbox{$X^2\Sigma^+(v^{\prime\prime}=1) \rightarrow B^2\Sigma^+(v^\prime=0)$}.  This includes the first spin-rotation and hyperfine-resolved laser spectroscopy of the MgF $B^2\Sigma ^+$($v=0$) state.
Furthermore, we demonstrate optical cycling in excess of 10 photons in the MgF molecule to efficiently optically pump molecules to vibrational states up to $v=2$.  Optical pumping enhances the spectroscopic signals in rovibrationally excited states with rotational quantum number $N=1$.

\section{Theory} \label{sec: Theory}

The $X^2\Sigma^+(v=0,1,2)$ and $B^2\Sigma^+(v=0)$ states are modeled using the following effective Hamiltonian:
\begin{equation}
\label{eq: Heff Sigma}
\begin{split}
    \hat{H}^{[^2\Sigma^+]}_{\mathrm{eff}} =& T_0 + B \textbf{N}^2 - D \textbf{N}^4 + \gamma \textbf{N} \cdot \textbf{S} \\
    &+ b_F \textbf{I} \cdot \textbf{S} + \frac{1}{3} c (3 I_z S_z - \textbf{I} \cdot \textbf{S}),
\end{split}
\end{equation}
where $\mathbf{S}$, $\mathbf{I}$, and $\mathbf{N}$ are the electron spin, fluorine nuclear spin, and rotational angular momenta, respectively.
This effective Hamiltonian accounts for the origin $T_0$, rotation $B$, centrifugal distortion $D$, electron spin-rotation $\gamma$ (from here on referred to as spin-rotation), and the Fermi-contact $b_F$ and dipole-dipole $c$ hyperfine interactions due to the fluorine nucleus ($I=1/2$). The $^{24}$Mg isotope has a nuclear spin $I=0$, and as we are concerned with the spectra of the $^{24}$MgF isotopologue here, all hyperfine parameters are the result of the fluorine nuclear spin. In the effective Hamiltionian given in \eqref{eq: Heff Sigma}, $T_0=T_e + G(v)$  accounts for both the electronic $T_e$ and vibrational $G(v)$ energy of the molecular state. Parameters for the $X^2\Sigma^+(v=0,1,2)$  states are taken from \cite{Anderson1994}\footnote{In the original microwave study the hyperfine parameters $b=b_F - c/3$ and $c$ are used}, while the parameters for the $B^2\Sigma^+(v=0)$  state determined by this study are summarized in Table \ref{tab:B state Results} and compared to prior work.

The  $A^2\Pi_{1/2}(v=0,1)$ states are modeled by the following effective Hamiltonian 
\begin{equation}\label{eq:Heff Pi}
    \begin{split}
        \hat{H}^{[^2\Pi_{1/2}]}_{\mathrm{eff}} =& T_0 +  B\textbf{J}^2 + a L_z I_z +  b_F \textbf{I} \cdot \textbf{S} \\
        &+ \frac{1}{3} c (3 I_z S_z - \textbf{I} \cdot \textbf{S}) -  \frac{1}{2} d (S_+ I_+ +S_- I_-) \\
        & \mp (-1)^{J-1/2}\frac{1}{2}(p+2q)(J+\frac{1}{2}),
    \end{split}
\end{equation}
where $L$ is the electron orbital angular momentum and $\mathbf{J}={\bf L}+{\bf S}$.
In addition to the parameters in \eqref{eq: Heff Sigma}, this effective Hamiltonian includes nuclear spin-orbit coupling $a$ and dipole-dipole coupling $d$. $\Lambda$-doubling is included phenomenologically by the final term of \eqref{eq:Heff Pi} with parameters $p$ and $q$, where the top (bottom) sign is taken for states of even (odd) parity. 
For simplicity, we will abbreviate $X^2\Sigma^+(v)$, $A^2\Pi_{1/2}(v)$, and $B^2\Sigma^+(v)$, as $X(v)$, $A(v)$, and $B(v)$ respectively in the remainder of this work.

Here, we are primarily concerned with identifying low rotational lines relevant to laser cooling.
For $X(0,1,2)\rightarrow A(0,1)$ transitions measured here, an insufficient number of levels were investigated to determine the $A(0,1)$ Hamiltonian parameters uniquely.
In our modeling, we use the $A(0)$ hyperfine parameters reported in Ref.\,\cite{Doppelbauer2022} for both the $v=0$ and $v=1$ states.
The expected few percent differences between $v=0$ and $v=1$ are within the uncertainty of the present work.
For the $X(1)\rightarrow B(0)$ spectrum, only the transitions originating from the low rotational states $N^{\prime\prime}\leq2$ of the $X(1)$ state and terminating in the $N^{\prime}\leq3$ levels of the $B(0)$ state or $J^\prime<5/2$ levels of the $A(0,1)$ states were observed and, therefore, the effective Hamiltonians were restricted to only include these levels.

We calculate all molecular Hamiltonians and dipole transition matrix elements using Python programming language via the {\tt pylcp} package \cite{Eckel2022}.
The effective Hamiltonians for the $X(v=0,1,2)$ and $B(0)$ states were constructed in a Hund's case (b$_{\beta J}$) basis. The energy levels for the $X(0,1,2)$ states were calculated by constructing and diagonalizing the $36\times36$ Hamiltonian for all magnetic sublevels of the $N^{\prime\prime}=0 - 2$ states. The energy levels of the $B(0)$ state were calculated by constructing and diagonalizing the $64\times64$ Hamiltonian for all magnetic sublevels of the $N^\prime =0-3$ states.   For the $A(0,1)$ states, the effective Hamiltonians were constructed in the Hund's case (a) basis and the energy levels were calculated by diagonalizing a $40\times40$ Hamiltonian for each parity and all magnetic sublevels of the $J^\prime = 1/2$ to $J^\prime = 7/2$ states.

\section{Experiment}

\subsection{Apparatus}
\begin{figure}
    \centering
    \includegraphics[width=\columnwidth]{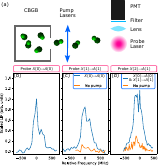}
    \caption{Optical pumping and fluorescence probe of a cold MgF beam. (a) Experimental schematic.  A molecular beam is formed using a cryogenic buffer gas beam (CBGB) source.  Fluorescence from the probe laser is detected on a photomultiplier (PMT) after passing through a lens assembly and bandpass interference filter.  One or more pump lasers are overlapped using a fiber optic beam combiner and quadruple passed for high pumping efficiency.  (b-d) Detected fluorescence signal as a function of probe laser frequency, with probe laser tuned near the $P_1(1)/Q_{12}(1)$ of the $X(0) \rightarrow A(0)$,
    $X(1) \rightarrow A(1)$, and
    $X(2) \rightarrow A(1)$ transitions, respectively.  In c and d, blue (orange) traces show the detected fluorescence with (without) pump lasers. All traces are normalized to the peak signal of (b).}
    \label{fig:optical_pumping_schematic}
\end{figure}

A schematic of the apparatus is shown in Fig.\,\ref{fig:optical_pumping_schematic}(a).
Our cryogenic buffer gas beam of MgF was previously described in Ref.\,\cite{Norrgard2023}. All aspects of the cryogenic buffer gas beam source are identical to the previous work except for the addition of a second stage buffer gas cell \cite{Lu2011,Hutzler2011b}. The second stage cell is made from high purity oxygen free copper, has a 9.53\,mm by 9.53\,mm square cross section, a length of 9.53\,mm, and a 6.25\,mm diameter circular aperture. The aperture is covered by a 30\,\% transparent copper mesh with an opening size of 0.15\,mm. There are two 1.59\,mm by 9.53\,mm vertically oriented rectangular slits in the sides of the second stage cell. The back of the second stage cell is completely open and is separated from the front aperture of the first stage cell by a 2.45\,mm gap.  

We orient our experiment by taking  $\hat{z}$ to  be the direction of travel of the molecular beam (roughly horizontal), $\hat{y}$ vertically upward, and $\hat{x}$ parallel to the ground and forming a right-handed coordinate system.  
The $X\rightarrow A$ transitions are  driven by three independent frequency-doubled Ti:sapphire lasers.  The $X\rightarrow B$ transitions are driven by a frequency-quadrupled fiber laser system. 
Fluorescence was detected along the $\hat{x}$ axis using a hybrid photomultiplier-avalanche photodiode.  Detection was made nearly free of background laser light by use of bandpass interference filters which pass only off-resonant vibronic fluorescence.

Laser light tuned near the transition under investigation is coupled to the fluorescence detection region by solarization-resistant multimode fibers.
The laser light is collimated to nominally 15\,mm 1/$e^2$ diameter and retroreflected such that it  propagates in the $\pm\hat{y}$-directions.
We observed noticeable solarization of the solarization-resistant fibers when transmitting 274\,nm laser light, thus we shutter the light prior to the fiber except during periods of fluorescence detection. 

For the $X\rightarrow A$ transitions, the absolute transition frequency was determined by monitoring a portion of the subharmonic light produced by the Ti:sapphire laser on a wavelength meter, which has a manufacturer stated 1-$\sigma$ uncertainty in the near-infared of 225~MHz.
Thus, the 1-$\sigma$ uncertainty on the $X\rightarrow A$ transitions is $550$\,MHz~\footnote{Unless states otherwise, all uncertainties in this paper are 1-$\sigma$.}.
For the $X\rightarrow B$ transition, a portion of the subharmonic 548\,nm light is sent to a saturated absorption spectroscopy iodine reference cell. 
We use the IodineSpec5 program \cite{knockel2004} to assign multiple iodine transitions near each MgF $X\rightarrow B$ transition.
The absolute accuracy of $X\rightarrow B$ transition frequencies is 14~MHz, limited by the statistical uncertainty in our fits of the MgF and I$_2$ line centers.

\subsection{Optical cycling and optical pumping}

\begin{figure*}
    \centering
    \includegraphics[width=\textwidth]{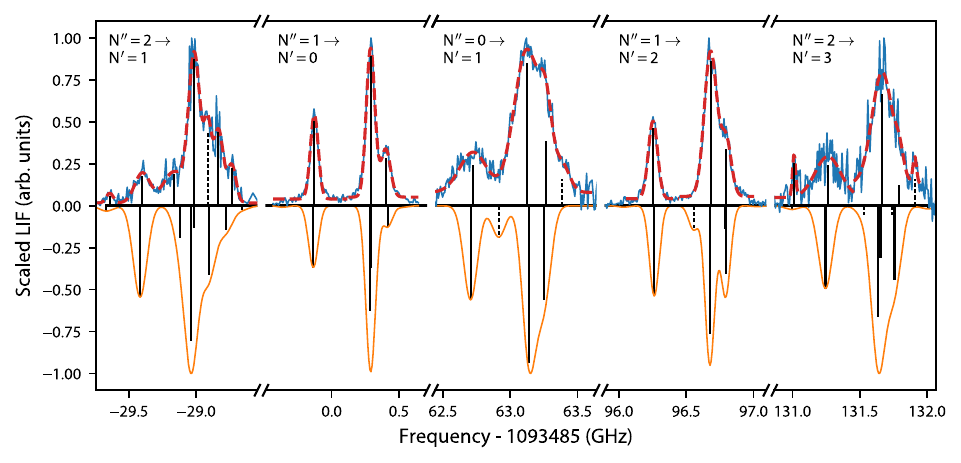}
    \caption
    {Observed (blue) and predicted Doppler (orange, reflected) and stick (black, reflected) spectrum of $X(1)\rightarrow B(0)$ transition. The observed spectra is fit to a sum of Gaussians (red dashed) to extract line centers (black). For the observed line centers and predicted stick spectrum, solid lines represent transitions that were assigned and dashed lines represent transitions that were not assigned or unresolved, respectively. 
    The observed LIF and predicted Doppler spectrum of each feature is normalized to its maximum.
    The predicted stick spectrum was scaled by the relative line strengths.
    Features from $N^{\prime\prime}=1$ have increased signal and decreased width due to optical pumping.
    }
    \label{fig:B state}
\end{figure*}
Optical pumping was used to increase the molecular population in the ground state of several transitions of interest.
A multimode fiber combiner is used to overlap up to three laser beams into a single output pumping beam.
The pumping beam intersects the molecular beam approximately 7\,cm from the output of the CBGB in a quadruple pass configuration. 
The pumping beam has a roughly 10\,mm 1/e$^2$ diameter on the first pass and roughly 30\,mm 1/e$^2$ diameter on the fourth pass due to residual divergence from the multimode fiber.

Figure \ref{fig:optical_pumping_schematic}(b-d) illustrates the sequential identification of the $P_1(1)$ lines of $X(0)\rightarrow A(0)$, $X(1)\rightarrow A(1)$, and $X(2)\rightarrow A(1)$.
We begin by tuning the probe laser near the $P_1(1)$ line of $X(0)\rightarrow A(0)$. 
After recording laser induced fluorescence (LIF) as a function of laser frequency, we tune a pump laser near to this line.  
To address all ground state spin-rotation and hyperfine components of the $P_1(1)$ line, pump lasers are phase modulated at 115\,MHz at modulation depth of approximatly 1.4~radians.  
With the probe tuned to the LIF maximum, the pump is scanned to optimize depletion of the LIF signal.  
We then tune the probe near the $X(1)\rightarrow A(1)$ $P_1(1)$ line.  
While a modest LIF signal (orange trace in Figure \ref{fig:optical_pumping_schematic}c) is detected without the $X(0) \rightarrow A(0)$ pump, the LIF signal is dramatically larger when optical pumping is applied (blue trace in Figure \ref{fig:optical_pumping_schematic}c).
The process is repeated to add a second $X(1) \rightarrow A(1)$ pump to aid in identifying the $X(2)\rightarrow A(1)$ $P_1(1)$ line.
Investigation of the $ X(v^{\prime\prime} = 1, N^{\prime\prime} = 1) \rightarrow B(v^\prime = 0, N^\prime)$ transitions are also assisted by the $X(0) \rightarrow A(0)$ pump.

Crudely, we estimate the optical pumping efficiency from $v=0$ by assuming each molecule scatters exactly one photon from the probe laser.  Scaling the fitted LIF amplitudes by the branching fractions of the detected transitions, we find the optical pumping efficiency of this setup is 60\,\% from $v=0$ to $v=1$, and 6\,\% from $v=0$ to $v=2$.
 We modeled  the optical pumping to $v=2$ with lasers $\mathcal{L}_{00}$ and $\mathcal{L}_{11}$ as a four-state ($v= 0 - 3)$ discrete Markov process.  Using the branching fractions measured in Ref.\,\cite{Norrgard2023}, we find that $N_\gamma \approx 60$ photon cycles are required to optically pump a fraction $(1-1/e)$ from initial state $v=0$ to final state $v=2$.  
 From these estimates, we find the average $v=0$ molecule cycles 10 photons in this simple model.



\subsection{B State spectroscopy}

The observed LIF spectrum of the $X\rightarrow B$ transition is presented in Fig. \ref{fig:B state}. A total of five groups of rotational branch features were observed: the $N^{\prime\prime}=2 \rightarrow N^{\prime}=1$, $N^{\prime\prime}=1 \rightarrow N^{\prime}=0$, $N^{\prime\prime}=0 \rightarrow N^{\prime}=1$, $N^{\prime\prime}=1 \rightarrow N^{\prime}=2$, and $N^{\prime\prime}=2 \rightarrow N^{\prime}=3$ features. 
The increased molecular population in the $X^2\Sigma^+(v=1,N=1)$ state provided by the optical pumping allowed the immediate assignment of the rotational quantum numbers of $N^{\prime\prime}=1 \rightarrow N^{\prime}=0$ and $N^{\prime\prime}=1 \rightarrow N^{\prime}=2$ features, from which the other rotational assignments were made. The optical pumping also provided increased LIF signal for the rotational features originating from the $N^{\prime\prime}=1$ levels. The observed frequencies of the rotational branch features are consistent with those reported in Ref.\,\cite{Barrow1967}.

The observed transition frequencies were determined via a nonlinear least-squares fit of the measured LIF spectrum to an appropriate number of Gaussian lineshapes \footnote{While a Voigt profile more accurately describes the combined contributions of the natural linewidth and Doppler broadening to the lineshape, the simpler Gaussian lineshape was used to fit for the line centers. Fitting with a Voigt profile makes no statistically significant difference in the fitted line center, due primarily to the symmetric nature of both lineshapes.}.
The amplitude, center, and width of each lineshape as well as the slope and intercept of a linear offset were floated in the fits.
The LIF data was cut so that a minimum number of spectral features were fit at a single time. The total number of Gaussian lineshapes included in a single fit was varied to optimize the fit. The Gaussian fits to the observed spectrum and the resulting centers of the Gaussian lineshapes are presented in Fig. \ref{fig:B state}. The line centers were used as the observed transition frequencies.
The fits indicate that the observed $N^{\prime\prime}=1 \rightarrow N^{\prime}=0$ and $N^{\prime\prime}=1 \rightarrow N^{\prime}=2$ branch features are narrower [85~MHz full width at half maximum (FWHM)] than the other observed features (125~MHz FWHM), likely due to molecules with lower transverse velocities being preferentially pumped by the optical fields. The finite spatial extent of the optical pumping beam along the probe axis results in molecules with lower transverse velocities along the probe axis to pass through the beam nearer to the center, where the laser intensity is higher. This provides increased optical pumping efficiency for these molecules.
In total, 22 individual features were observed.
Of these, 19 were uniquely assigned to 24 transitions using an iterative process described in Appendix~\ref{sec:appendix_B_state}.
The frequencies of the observed features and the relevant assignments are presented in Table \ref{tab:B state lines/assignments}.

The 24 assigned transition frequencies were fit using non-linear least squares to the effective Hamiltonian \eqref{eq: Heff Sigma} with fit parameters $T_0$, $B$, $\gamma$, $b_F$, and $c$.
Because we only probe low $J$, we fix the value of $D$ to the value from a previous high-temperature study by Barrow and Beale\,\cite{Barrow1967}\,\cite{Barrow1967}.
In the fit, each transition was weighted by the inverse of the uncertainty in the transition frequency that resulted from the Gaussian lineshape fits.
The fit resulted in a root mean square (RMS) of the weighted residuals of 25.5~MHz, commensurate with the measurement uncertainty of 14~MHz.
The difference between the calculated and measured transition frequencies are presented in Appendix~\ref{sec:appendix_B_state}, Table \ref{tab:B state lines/assignments}.
The resulting predicted Doppler and stick spectrum generated using the optimized parameters from the fit are shown in Fig.~\ref{fig:B state}. The spectral predictions were made following the procedure described in Appendix~\ref{sec:appendix_B_state}.

Our parameters for the $B(v=0)$ state Hamiltonian are presented in Table \ref{tab:B state Results}, together with those of Ref.~\cite{Barrow1967}.
The rotational constants differ by 1.3~MHz and are not consistent at the estimated $1$-$\sigma$ level.
This discrepancy is not surprising as the two studies probed different $J$ levels of the $B(0)$ state.
The previous high temperature study did not have high enough resolution to observe the fine and hyperfine structure.

\begin{figure}
    \centering
    \includegraphics[width=\columnwidth]{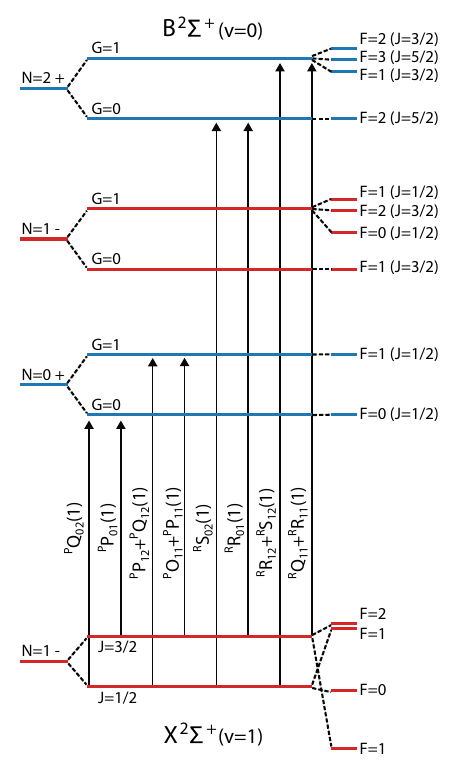}
    \caption
    {Energy levels and associated quantum numbers of the $N= 0,1,$ and $2$ rotational levels of the $B(0)$ state and the $N=1$ rotational level of the $X(1)$ state.
    Even parity states are blue; odd parity states are red.
    A transition from each of the eight individual branches is shown as an example.
    While not well defined, the approximate value of $J$ for each hyperfine level of the $B(0)$ state is shown to motivate the the $\Delta J$ component of each branch assignment.
    }
    \label{fig:XtoB levels}
\end{figure}

Figure \ref{fig:XtoB levels} shows the resulting schematic of the  $N^\prime =0$, 1, and 2 levels of the $B(0)$ state with the $N^{\prime\prime}=1$ levels of the $X(1)$ state.
In Fig. \ref{fig:XtoB levels}, the spacing between the hyperfine levels within each rotational state are drawn to scale.
The $B(0)$ state exhibits the pattern typical for Hund's case(b)$_{(\beta S)}$, which occurs when $b_F\gg\gamma$~\footnote{
Note that the energy level pattern of the $N^{\prime\prime}=1$ level of the $X(1)$ state does not resemble that of a typical Hund's case(b$_{\beta J}$) state but is in a more intermediate regime between a Hund's case(b$_{\beta J}$) and Hund's case(b$_{\beta S}$).
}.
In this scenario, \eqref{eq: Heff Sigma} reveals that $\textbf{S}$ first couples to $\textbf{I}$, resulting in the approximately good intermediate angular momentum $\textbf{G} = \textbf{S}+\textbf{I}$.
$\textbf{G}$ then couples to the $\textbf{N}$ to form the total angular momentum $\textbf{F} = \textbf{G} + \textbf{N}$.
This results in each rotational level of the $B^2\Sigma^+$ state being separated into two subsets of levels defined by $G$: $G=I-S=0$ and $G=I+S=1$.


The hyperfine structure of the $B(0)$ state manifests in the structure of each rotational branch feature as a set of lower frequency transitions to the $G=0$ level and a set of higher frequency transitions to the $G=1$ levels as seen by the two, well-separated peaks for each branch shown in Fig.~\ref{fig:B state}.
Here, we use a modified version of the typical Hund's case(b$_{\beta J})$ $\rightarrow$ Hund's case(b$_{\beta J})$ branch designation scheme \cite{steimle1997}, $^{\Delta N}\Delta J_{F_i,F_j}(N^{\prime\prime})$, where the first $F_i$ designation is replaced with the intermediately good quantum number $G=I+S=0,1$ of the $B(0)$ state and the second $F_j$ designation takes the typical value ($F_j=1$ for $J''=N^{\prime\prime}+S''$ and $F_j=2$ for $J''=N^{\prime\prime}-S''$) for the $X(1)$ state. According to this branch designation, the observed $X(1)\rightarrow B(0)$ transition has eight separate branches: $^PQ_{02}$, $^PP_{01}$, $^PP_{12} + ^PQ_{12}$, $^PO_{11} + ^PP_{11}$, $^RS_{02}$, $^RR_{01}$, $^RR_{12} + ^RS_{12}$, and $^RQ_{11} + ^RR_{11}$. Each of these branches are shown in Fig. \ref{fig:XtoB levels}. Note that $J'$ is not a good quantum number in the $B(0)$ state, and therefore, the $\Delta J$ value given in each branch designation is approximate.   



\begin{table}
    \centering
    \caption{Best fit parameters of the $B^2\Sigma^+(v=0)$ state of MgF, with comparison to prior work. All values are in MHz.}
    \begin{tabularx}{\columnwidth}{cdd}
    \hline\hline
    Parameter   & \text{This work}\footnotemark[1]& \text{Ref.\,\cite{Barrow1967}}\footnotemark[1]   \\
    \hline
$T_0$       & 1\,114\,851\,907.(3) & 1\,114\,852\,000.(300)              \\
 $B$         & 16\,066.9(8)      & 16\,065.6(3)                   \\
 $D$         & 0.032\,47\footnotemark[2]       & 0.032\,47(3)                   \\
 $\gamma$    & 21.(6)         &                              \\
 $b_F$       & 443.(18)       &                              \\
 $c$         & 653.(56)       &                              \\
    \hline\hline
    \end{tabularx}
    \footnotetext[1]{Values in parenthesis are the 1-$\sigma$ standard error determined from the fit.}
    \footnotetext[2]{Fixed to value in Ref.\,\cite{Barrow1967}.}
    \label{tab:B state Results}
\end{table}


The fitted hyperfine parameters of the $B(0)$ state are a few times larger than those of the $X(0)$ state.
The fitted $b_F$ is 2.1 times larger than that of the $X(0)$, while the dipolar $c$ parameter is 3.7 times larger~\cite{Anderson1994}, indicating that the electron density at the fluorine nucleus is larger in the $B(0)$ state than in the $X(0)$ state.
Additionally, while the ratio $c/b_F=1.5$ in the $B(0)$ state, in the  $X(0)$ state this ratio is only $c/b_F = 0.8$.
The much larger $c$ parameter in the $B(0)$ state indicates that the electron distribution around the fluorine nucleus is more anisotropic than in the $X(0)$ state. Finally, the fluorine hyperfine interactions in the $B(0)$ state of MgF are much larger than that of the other alkaline-earth monofluorides CaF and SrF, where the fluorine hyperfine structure is unresolved~\cite{Devlin2015,steimle1977}. This is similar to the $A(0)$ state of these three molecules, where the $A(0)$ state of MgF has resolved fluorine hyperfine structure and the $A(0)$ states of CaF and SrF do not~\cite{Doppelbauer2022}. These larger fluorine hyperfine interactions provide additional support for the conclusion that the bonding in MgF is more covalent than that of the other alkaline-earth monofluorides~\cite{Anderson1994,Doppelbauer2022}. 

The fitted spin-rotation parameter $\gamma$ in the $B(0)$ state is roughly 41~\% of the $X(0)$ state value~\cite{Anderson1994}.
This is in contrast to the the other alkaline-earth metal fluorides CaF and SrF, where the spin-rotation parameter of the $B(0)$ state is larger in magnitude and opposite in sign when compared to the ground $X(0)$ state.
For CaF, the ratio of $\gamma$ of the $B(0)$ state to that of the ground $X(0)$ state is -35.0; for SrF, this ratio is $-33.5$~\cite{Devlin2015,steimle1977}.

For CaF and SrF, the $\gamma$ of the $B(0)$ state are $-1378.87$~MHz and $-4020$~MHz, respectively, which are both larger and have the opposite sign compared to our fitted value of $\gamma $ for $B(0)$ state of MgF, 21~MHz.
Because of the value of $\gamma$ in a $^2\Sigma^+$ state depends on  second order effects which mix the $^2\Sigma^+$ state with nearby $^2\Pi_{1/2}$ states~\cite{Brown2003,Lim2017}, it is perhaps unsurprising that the value of $\gamma$ for MgF is so strikingly different.
The energy spacing of neighboring electronic levels is roughly a factor of 4 larger for MgF.
Moreover, an atomic d state contributes to the electronic structure of the $A^2\Pi$ and $B^2\Sigma^+$ states of CaF and SrF, but is absent in MgF~\cite{NIST_ASD_2023}.
Better determination of spectroscopic constants, particularly of the $^{25}$MgF isotopologue, can provide insight into this value of $\gamma$.

The primary goal of the $X(1)\rightarrow B(0)$ spectroscopy was to identify the rotationally closed vibrational repumping transitions at high resolution. 
Due to parity selection rules, rotational closure can be ensured by driving transitions from all sublevels of the $X(v^{\prime\prime}, N^{\prime\prime}=1)$ state to the same subset of levels in the $B(v^\prime, N^\prime = 0)$ state. For example, simultaneously driving the $^PP_{12}+^PQ_{12}(1)$ transitions and the $^PO_{11}+^PP_{11}(1)$ transitions will effectively repump any decays to $X(1)$  during laser cooling. Repumping through the $B(0)$ state as opposed to the $A(0)$ state avoids creating a $\Lambda$ system and keeps the scattering rate for the main cycling transition high. Repumping through the $B(0)$ state can lead to loss of rotational closure through $ B \rightarrow A \rightarrow X$ cascade decays, but this should only occur at the $1.2\times 10^{-4}$ level \cite{Norrgard2023}.

\subsection{A state spectroscopy}

\begin{figure}
    \centering
    \includegraphics[width=1.1\columnwidth]{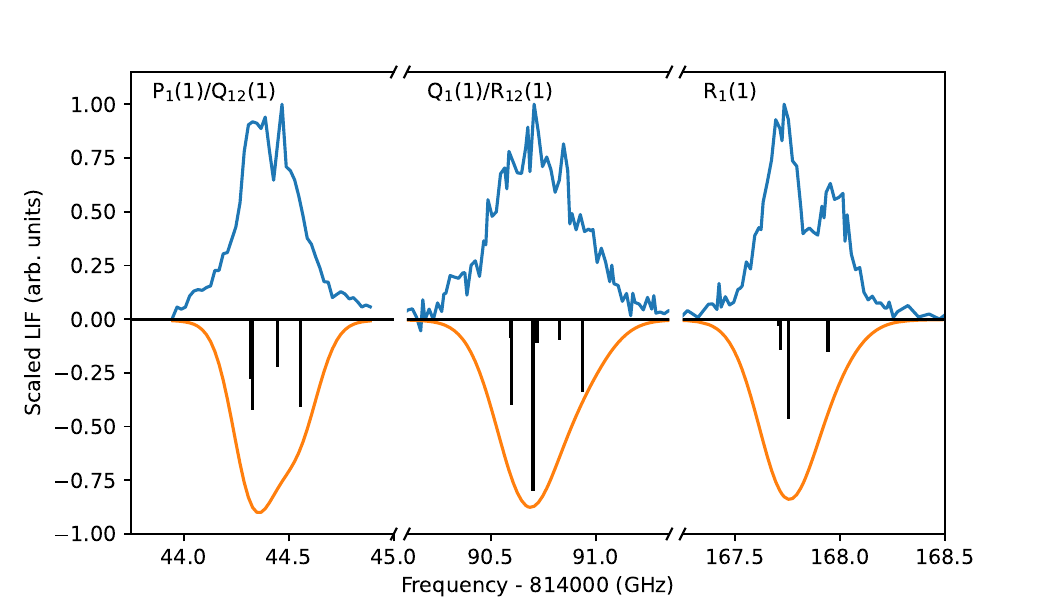}
    \caption
    {Observed (blue) and predicted Doppler (orange, reflected) and stick (black, reflected) spectrum of $X(2)\rightarrow A(1)$ transition.
    The observed LIF and predicted Doppler spectrum of each feature is normalized to its maximum.
    The predicted stick spectrum was scaled by the relative line strengths.
    }
    \label{fig:XA v21}
\end{figure}

\begin{table}[b]
    \centering
    \caption{MgF transitions for optical cycling.  Numbers in parenthesis represent the 1-$\sigma$ uncertainty.}
    \begin{tabular}{cdc}
    \hline\hline
        Label               & \multicolumn{1}{c}{Observed (GHz)} & Reference \\
        \hline 
        
        \vspace{6 pt}
        $X(0)\rightarrow A(0)$\footnotemark[1]     
                            & 834\,294.32(55)  & This work \\
                            & 834\,294.356(10) & 
                            \cite{Doppelbauer2022} \\
                            & 834\,294.356(30) & \cite{Gu2022erratum}\\
                            
                            & 834\,291.11(11) & \cite{Barrow1967}\\
         &  & \\
        $X(1)\rightarrow A(1)$\footnotemark[1]     
                            & 835\,128.48(55)  & This work \\
                           
         &  & \\

         $X(1)\rightarrow A(0)$\footnotemark[1]  
                            & 812\,959.14(55)& This work \\
                            & 812\,959.246(30) & \cite{Xu2019b}\\
                            & 812\,955.46(11) & \cite{Barrow1967}\\
         \\                 
         
        $X(1)\rightarrow B(0)$\footnotemark[2]  
                           &1\,093\,485.289(14)  & This work\\
        $X(1)\rightarrow B(0)$\footnotemark[3] &1\,093\,485.402(14)  & This work\\
                            &1\,093\,485.07(9) & \cite{Barrow1967}\\

         &  & \\
         $X(2)\rightarrow A(1)$\footnotemark[1] & 814\,044.49(55) & This work \\
         &  & \\
         &  & \\
         \hline\hline
    \end{tabular}
    \footnotetext[1]{$P_1(1)$/$Q_{12}(1)$ branch features.}
    \footnotetext[2]{$^PP_{12}+^PQ_{12}(1)$/$^PO_{11}+^PP_{11}(1)$ branch features.}
    \footnotetext[3]{$^PP_{12}+^PQ_{12}(1)$ branch feature.}
    \label{tab:cycling transitions}
\end{table}

A total of six spectral features from the $X(0)\rightarrow A(0)$, $X(1)\rightarrow A(1)$, $X(1)\rightarrow A(0)$, and $X(2)\rightarrow A(1)$ transitions were observed.
The observed transition frequencies are reported in Appendix~\ref{sec:appendix_A_state}, Table \ref{tab:All A state lines}. 
The uncertainty in the absolute frequency of each spectral feature is larger than the observed width, and, therefore, only the absolute frequency of the peak of the observed LIF signal is reported.
All observed $X\rightarrow A$ transitions originate from an $N^{\prime\prime}=1$ level of the $X(0)$, $X(1)$, or $X(2)$ state.
For each of the $X(0)\rightarrow A(0)$, $X(1)\rightarrow A(1)$, and $X(1)\rightarrow A(0)$ transitions a single blended spectral feature corresponding to the combination of the $P_1(1)$ and $Q_{12}(1)$ features was observed.
The observed $P_1(1)$/$Q_{12}(1)$ features of the $X(0)\rightarrow A(0)$ and $X(1)\rightarrow A(1)$ transitions are show in Fig. \ref{fig:optical_pumping_schematic}(b) and (c) respectively.

For the $X(2)\rightarrow A(1)$ transition, a total of 3 spectral features were observed: the blended $P_1(1)$/$Q_{12}(1)$, $Q_1(1)$/$R_{12}(1)$, and $R_1(1)$ features.
The observed LIF spectrum of these three features is presented in Fig. \ref{fig:XA v21} together with the predicted stick and Doppler spectrum.
The $X(v=0,1,2)$ state parameters were fixed to the values of Ref.~\cite{Anderson1994} and the $A(v=0,1)$ state hyperfine parameters were fixed to the $A(0)$ values of Ref.~\cite{Doppelbauer2022}. 
Each rotational transition was predicted separately to remove the dependence on the rotational and $\Lambda$-doubling parameters of the $A(1)$ state.
The overall linecenter and width of each rotational transition was determined via a nonlinear least-squares fit of the predicted Doppler spectrum to the LIF signal.

Comparing the frequencies of the three features of Fig.~\ref{fig:XA v21}, we find the A$^2\Pi_{1/2}(v=1)$ rotational constant to be $B=15430(40)$\,MHz, with the uncertainty dominated by uncertainty in the relative frequency of the laser.
We also find the linear combination of $\Lambda$-doubling parameters $p+2q$ to be consistent with zero within our 40\,MHz uncertainty; this is expected as these values should not substantially differ from the  A$^2\Pi_{1/2}(v=0)$ values of  $p+2q = 15$\,MHz~\cite{Doppelbauer2022}.

\section{conclusion}

Figure \ref{fig:BS} presents two possible laser cooling schemes for MgF which can cycle up to to roughly $2\times 10^4$ photons, sufficient for laser cooling and magneto-optically trapping.  In this work, we have identified several $X\rightarrow A$ and $X\rightarrow B$ transitions in MgF, including all rotationally closed transitions for laser cooling depicted in Fig.\,\ref{fig:BS}.  We summarize the rotationally closed transitions in Table~\ref{tab:cycling transitions}, with additional assigned transitions tabulated in Appendices \ref{sec:appendix_A_state} and \ref{sec:appendix_B_state}. It is important to note that in either laser cooling scheme stray electric fields will need to be controlled to below 1~V/cm in order to prevent loss of rotational closure at the 10$^4$ level \cite{Doppelbauer2022}. This level of control can be easily achieved by grounding the vacuum chamber. 
Additionally, electric field-induced rotational branching will result in decays to the $N=0, 2$ levels of the $X(0)$ state, the same levels populated by magnetic dipole decays from $A(0)$ and $B(0)$ \cite{DiRosa2004}, or by electric dipole cascade decays from $B(0)$ state through the intermediate $A(0)$ state. Losses to these states can be easily repumped, e.g.\ with microwaves \cite{Norrgard2016, NorrgardThesis}.

Utilizing the $X(0)\rightarrow A(0)$ and $X(1)\rightarrow A(1)$ transitions, we optically pumped molecules into the $N=1$ level of the $X(1)$ and $X(2)$ states, cycling on average 10 photons.
The increased molecular population in the excited vibrational levels provided increased signal when detecting the vibrational repumping transitions.
In addition to identifying all needed vibrational repumping transitions for MgF, the observed spectra of the $X(1)\rightarrow B(0)$ allowed the spin-rotation and hyperfine parameters of the $B(0)$ state to be determined.
The data and observations in this work lay the necessary foundation for laser cooling and trapping MgF.

\section*{Acknowledgments}
The authors thank Timothy Steimle for advice on the assignment and fitting of the $X^2\Sigma^+(1) \rightarrow B^2\Sigma^+(0)$ data. The authors thank Jacek Klos, Daniel Barker, and Weston Tew for a careful reading of the manuscript. NHP was supported by the NRC postdoctoral fellowship. This work was supported by NIST.


\newpage
\appendix
\section{$X\rightarrow A$ State Transition frequencies}


Table~\ref{tab:All A state lines} contains all frequencies for the $X\rightarrow A$ state transitions observed in this work.
Note that these frequencies correspond to the maximum LIF signal for each feature.

\label{sec:appendix_A_state}
\begin{table}[h]
    \centering
    \caption{Observed frequencies and assignments of the $X^2\Sigma^+(v=0,1,2)\rightarrow A^2\Pi_{1/2}(v=0,1)$ transitions of MgF. The vibrational and rotational quantum numbers of the ground and excited states are presented. Numbers in parenthesis represent the 1-$\sigma$ uncertainty.}
    \label{tab:All A state lines}
    \begin{tabular}{cccd cc}
    \hline\hline
       $v^{\prime\prime}$ & $v^\prime$ & Transition &\multicolumn{1}{c}{Observed (GHz)} & $J^\prime$  & $N^{\prime\prime}$ \\
    \hline
        0 & 0 & $P_1(1)$/$Q_{12}(1)$ & 834\,294.32(55) & 0.5 & 1 \\
        1 & 1 & $P_1(1)$/$Q_{12}(1)$ & 835\,128.48(55) & 0.5 & 1 \\
        1 & 0 & $P_1(1)$/$Q_{12}(1)$ & 812\,959.14(55) & 0.5 & 1 \\
        2 & 1 & $P_1(1)$/$Q_{12}(1)$ & 814\,044.49(55) & 0.5 & 1 \\
        2 & 1 & $Q_1(1)$/$R_{12}(1)$ & 814\,090.66(55)  & 1.5 & 1  \\
        2 & 1 & $R_1(1)$ & 814\,167.60(55)  & 2.5 & 1  \\
    \hline\hline
    \end{tabular}
    \label{tab:garbage}
\end{table}


\FloatBarrier 

\section{$X\rightarrow B$ transition frequencies, assignments, and fits}
\label{sec:appendix_B_state}
Table~\ref{tab:B state lines/assignments} contains all frequencies for the $X(1)\rightarrow B(0)$ transitions and their assignments observed in this work. In addition to the 24 assigned transitions, three other unassigned transitions were observed. These transitions may be due to the other, less abundant, isotopologues of MgF, $^{25}$MgF and $^{26}$MgF.

Spectral assignments were made via an iterative process. Initial assignments of the fine and hyperfine components of each rotational branch were made using both combination differences and spectral predictions. Combination differences using the known $X(1)$ state energy levels \cite{Anderson1994} were used to assign the $^PP_{01}(1)$, $^PO_{11}+^PP_{11}(1)$, and $^PP_{12}+^PQ_{12}(1)$ features and determine the energy of the $N=0$, $G=0$, $F=0$ and $N=0$, $G=1$, $F=1$ levels of the $B(0)$ state.
Assuming that these levels are split by the Fermi-contact interaction, an initial estimate for the Fermi-contact parameter of $b_F=655$ MHz was made. 
Using this estimated $b_F$ value, a prediction of the $X(1) \rightarrow B(0)$ spectrum was then made using the effective Hamiltonian \eqref{eq: Heff Sigma}  with initial rotational parameters for the $B(0)$ state set to those of Ref.\,\cite{Barrow1967}. The spin rotation parameter $\gamma$ was initially set to zero and the dipolar magnetic hyperfine parameter $c$ was set to 10 MHz to break the degeneracy between the three $F$ sublevels of each $G=1$ manifold. Additional initial assignments were made with the aid of this spectral prediction. These assignments were confirmed to match the assignments of several features made via the combination differences. This procedure resulted in the initial assignment of 14 transitions to 14 unique spectral features.

These 14 spectral assignments were then used as inputs to an unweighted nonlinear least-squares fit to the transition frequencies using the effective Hamiltionian \eqref{eq: Heff Sigma}. The parameters of the $X(1)$ state and the centrifugal distortion parameter of the $B(0)$ state, $D'$, were held fixed and the Origin, $T_0'$, rotational, $B'$, spin-rotation, $\gamma'$, and the magnetic hyperfine parameters, $b_F'$ and $c'$, of the $B(0)$ state were varied in the fit. Each transition was weighted equally to avoid errors from a possible incorrect assignment of a transition. The results from this fit were then used to produce another spectral prediction which was used to make additional assignments.
These additional assignments were made with the restriction of assigning, at most, only a single transition to each observed spectral feature. This resulted in the assignment of an additional 5 transitions to 5 unique spectral features.
This expanded set of 19 transitions were then used as inputs to the unweighted nonlinear least-squares fitting algorithm.
The same set of parameters were held fixed and varied as in the previous fit and all transitions were again weighted equally.
The results of this second fit were then used to produce a final spectral prediction from which the last set of assignments were made.
In total, this iterative procedure resulted in the assignment of 24 transitions to 19 unique spectral features. These 24 transitions were used as inputs to the final weighted nonlinear least-squares fit to the effective Hamiltonian \eqref{eq: Heff Sigma} described in the text. The assigned transition frequencies along with the branch designations and associated quantum numbers are presented in Table \ref{tab:B state lines/assignments}.

Spectral predictions were made in the following manner: First, the relative transitions amplitudes are determined by rotating the transition dipole matrix (TDM) for each vibronic transition into the free-field eigen basis. Then each non-zero transition amplitude is associated with the appropriate transition's frequency, the difference between the energies of the associated ground and excited states. The ground and excited state energies were determined via the diagonalization of the appropriate effective Hamiltonian described in Sec. \ref{sec: Theory}. To account for the degeneracy of the $M_F$ levels, the amplitudes of all degenerate transitions between sets of states who only differ by the degenerate $M_F$ quantum numbers are summed to get the total transition amplitude. This set of corresponding transition frequencies and relative total transition amplitudes produces the simulated stick spectrum. The simulated Doppler spectrum is produced from the stick spectrum by providing each predicted transition a Gaussian lineshape.

\begin{table}
    \centering
    \caption{Observed transition frequencies in MHz and assignments for the $X^2\Sigma^+(1)\rightarrow B^2\Sigma^+(0)$ band of MgF. Also presented are the differences between the observed (Obs.) and calculated (Calc.) transition  frequencies. The calculated transition frequencies where obtained using the optimized parameters determined from the fit of the $X^2\Sigma^+(1)\rightarrow B^2\Sigma^+(0)$ data. The fit resulted in an RMS of the residuals of 25.5\,MHz. The absolute accuracy of the observed transition frequencies is 14~MHz, limited by the statistical uncertainty in our fits of the MgF and I$_2$ line centers}
    \label{tab:B state lines/assignments}
   \begin{tabular}{lcccccccc}
\hline\hline
 Branch              & $N^{\prime\prime}$   & $J^{\prime\prime}$   & $F^{\prime\prime}$   & $N^\prime$   & $G^\prime$   & $F^\prime$   & Obs.           & Obs. $-$ Calc.   \\
\hline
 $^PQ_{02}$          & 2                    & 1.5                  & 2                    & 1            & 0            & 1            & 1\,093\,455\,600     & 17.0             \\
 $^PP_{01}$          & 1                    & 1.5                  & 1                    & 0            & 0            & 0            & 1\,093\,484\,872     & 8.0              \\
           & 2                    & 2.5                  & 2                    & 1            & 0            & 1            & 1\,093\,455\,358     & 25.0             \\
  $^PP_{12}+^PQ_{12}$ & 1                    & 0.5                  & 1                    & 0            & 1            & 1            & 1\,093\,485\,289     & -7.0             \\
  & 1                    & 0.5                  & 0                    & 0            & 1            & 1            & 1\,093\,485\,402     & -13.0            \\
  & 2                    & 1.5                  & 1                    & 1            & 1            & 0            & 1\,093\,455\,834     & -47.0            \\
  & 2                    & 1.5                  & 1                    & 1            & 1            & 1            & 1\,093\,456\,264     & 46.0             \\
 $^PO_{11}+^PP_{11}$ & 1                    & 1.5                  & 2                    & 0            & 1            & 1            & 1\,093\,485\,289     & 2.0              \\
  & 2                    & 2.5                  & 3                    & 1            & 1            & 2            & 1\,093\,455\,986     & 24.0             \\
  & 2                    & 2.5                  & 2                    & 1            & 1            & 2            & 1\,093\,455\,986     & 3.0              \\
  & 2                    & 2.5                  & 2                    & 1            & 1            & 1            & 1\,093\,456\,165     & 72.0             \\
 $^RR_{01}$          & 0                    & 0.5                  & 0                    & 1            & 0            & 1            & 1\,093\,547\,724     & 19.0             \\
           & 1                    & 1.5                  & 1                    & 2            & 0            & 2            & 1\,093\,581\,253     & -10.0            \\
           & 2                    & 2.5                  & 2                    & 3            & 0            & 3            & 1\,093\,616\,012     & 16.0              \\
 $^RQ_{11}+^RR_{11}$ & 0                    & 0.5                  & 1                    & 1            & 1            & 2            & 1\,093\,548\,123     & -18.0            \\
  & 0                    & 0.5                  & 1                    & 1            & 1            & 1            & 1\,093\,548\,264     & 13.0              \\
  & 1                    & 1.5                  & 2                    & 2            & 1            & 3            & 1\,093\,581\,681     & 4.0              \\
  & 1                    & 1.5                  & 2                    & 2            & 1            & 2            & 1\,093\,581\,797     & 11.0             \\
  & 2                    & 2.5                  & 3                    & 3            & 1            & 4            & 1\,093\,616\,667     & 34.0             \\
  & 2                    & 2.5                  & 2                    & 3            & 1            & 3            & 1\,093\,616\,795     & 37.0             \\
 $^RR_{12}+^RS_{12}$ & 1                    & 0.5                  & 0                    & 2            & 1            & 1            & 1\,093\,581\,681     & 6.0              \\
  & 1                    & 0.5                  & 1                    & 2            & 1            & 2            & 1\,093\,581\,797     & 2.0            \\
  & 2                    & 1.5                  & 1                    & 3            & 1            & 2            & 1\,093\,616\,667     & 13.0              \\
  $^RS_{02}$          & 2                    & 1.5                  & 2                    & 3            & 0            & 3            & 1\,093\,616\,263     & 17.0              \\
                      Unassigned\footnotemark[1] &          &                      &                      &              &              &              & 1\,093\,456\,085 &                  \\
                      Unassigned &          &                      &                      &              &              &              & 1\,093\,548\,384     &                  \\
                      Unassigned &          &                      &                      &              &              &              & 1\,093\,616\,909     &                  \\
\hline\hline
\end{tabular}
\footnotetext[1]{This peak was fit to a separate Gaussian in the analysis but may not be a separate feature from the neighboring peak at 1093456165\,MHz.}
\end{table}

\FloatBarrier

\bibliography{thebib}

\end{document}